# Simultanous two-wavelength phase unwrapping using external module for multiplexing off-axis holography

Nir A. Turko, Natan T. Shaked[*]

*Department of Biomedical Engineering, Faculty of Engineering, Tel-Aviv University, Tel-Aviv 69978, Israel*
[*]*Corresponding author: nshaked@tau.ac.il*

**Abstract:** We present a dual-wavelength external holographic microscopy module for quantitative phase imaging of 3D structures with extended thickness range. This is done by simultaneous acquisition of two off-axis interferograms, each of which at a different wavelength, and generation of a synthetic wavelength, which is larger than the sample optical thickness, allowing two-wavelength unwrapping. The simultaneous acquisition is carried out by using optical multiplexing of the two interferograms onto the camera, where each of them has orthogonal off-axis interference fringe direction in relation to the other one. We used the system to quantitatively image a 7.96 μm step target and 30.5 μm circular copper pillars.

http://dx.doi.org/10.1364/OL.99.099999

Interferometric phase microscopy (IPM), also called digital holographic microscopy, provides quantitative optical thickness measurements for biological studies [1, 2] and high-accuracy profiling in metrology and surface inspection [3]. In this technique, the quantitative phase of the light interacted with the sample is reconstructed from an inteferogram of the sample [4]. The phase acquired is proportional to the surface topography for a reflective sample, or to the integral refractive index for a transparent sample. Since the phase of light is $2\pi$-periodic, objects, which are optically thicker than the illumination wavelength, are wrapped, and subject to phase measurement ambiguity. A 2D phase unwrapping algorithm can be digitally applied to obtain continuous phase reconstruction [5]. These algorithms, however, have two major drawbacks; they are computationally-demanding, and they fail when a large phase discontinuity is encountered, such as in steep steps or sharp refractive-index variations.

Alternatively, a system-based solution to the phase ambiguity problem is two-wavelength interferometry [6]. Using this technique, two interferograms with different illumination wavelengths are acquired, and the wrapped phase profile is extracted from each of them separately. Then, by simple processing, as described later, a new phase map with a large synthetic wavelength is obtained, significantly increasing the unambiguous phase range [7–9]. Since two different interferograms are needed per each sample instance, the acquisition should be faster than sample dynamics. A more general solution is simultaneous dual-wavelength interferometric acquisition.

In 2007, Kuhn et al. [8] used simultaneous two-wavelength holography by multiplexing two beams of different wavelengths on the same sensor, providing real-time holographic capabilities. Their system was based on two Mach-Zehnder interferometers, built around the sample, creating two separate reference beam paths, one for each wavelength, so that on the camera they obtained two off-axis interferograms of the sample simultaneously with 90°-rotated fringe direction in relation to each other. Other methods for separating the two wavelengths in simultaneous dual-wavelength interferometry include using a color Bayer-mosaic camera [10], and using polarization [11]. Variants of the techniques described above were used for examination of samples with large topography changes, such as porous coal samples [12], and biological samples [13]. However, all of these methods require two separate reference beams, which are independent of the sample on most of the optical path. These setups are prone to mechanical noise, as all three beams may not be subjected to the same vibrations. Specifically, when creating a large synthetic wavelength, the result is sensitive to noise even more [7].

In parallel to dual-wavelength interferometry, self-interference interferometric techniques evolved over the past decade [14-16], allowing more stable systems with decreased temporal phase noise due to nearly common-path interferometric geometry. In these systems, both the reference and the sample beams are formed from the image of the sample itself. The reference beam can be generated externally, after the output image plane of the optical system, from a spatially filtered version of the image, effectively erasing the sample spatial modulation from one of the beams, while the off-axis interference is realized by a retro-reflector [15] or a diffraction grating [16]. These setups not only increase the temporal stability of the system, but also make it less complex and more portable and compact, since they are external to the imaging system and not built around the sample.

Recently, we presented an external interferometric module design to allow doubling of the imaged field of view [17]. We used two beam splitters and two retro-reflectors to create a multiplexed off-axis interferogram, containing two fields of view of the sample at once, each of which was encoded into another interference fringes direction. Since the fringe directions were orthogonal, both fields of view could be retrieved; thus, doubling the imaged area of the sample, while sharing the dynamic range of the camera in the acquired multiplexed hologram. This idea was lately extended to multiplexing interferometric phase image and interferometric fluorescence image with simultaneous acquisition [18].

In the current letter, we present two-wavelength phase unwrapping using an external dual-wavelength interferometric module, which is based on off-axis interferometric multiplexing. As presented in Fig. 1(a), we used a reflectance microscope as the imaging setup, and to its output we connected the proposed dual-wavelength module, illustrated in Fig. 1(b). The imaging system was illuminated by a supercontinuum source (SC400-4; Fianium), connected into an acousto-optical tunable filter (SC–AOTF, Fianium), which created two simultaneous spectral bandwidths ($\lambda_1$=580 nm, $\lambda_2$=597 nm, or alternatively $\lambda_1$=580 nm, $\lambda_2$=605 nm, all with spectral bandwidth of approximately 5.4 nm). The illumination beam was first expanded by a beam expander (lenses L1, $f$ = 50 mm and L2, $f$ = 400 mm), which was followed by a 4-f lens configuration (lenses L3, $f$ = 200 mm, beam splitter BS1, and MO, 20×, 0.4 NA). After being reflected from the sample, the same microscope objective magnified the image and projected it through a tube lens (L4, $f$ = 200 mm) on the image plane of the microscope, where we attached the dual-wavelength multiplexed interferometric module.

As shown in Fig. 1(b), in this module the beam was Fourier transformed by lens L5 ($f$ = 160 mm), while being split into sample and reference beams by beam splitter BS2. The sample beam was Fourier transformed by lens L6 ($f$ = 75 mm), reflected by mirror M1, and projected into the monochrome camera (DCC1545M, Thorlabs) through lenses L6 and L8 ($f$ = 200 mm), so that the image-plane amplitude and phase were projected onto the camera from both wavelength channels at once.

In the reference beam arm of BS2, at the Fourier plane of lens L5, we positioned a spatial filter made out of a 30 μm pinhole PH, which selected only the low-frequency spatial information of the image and effectively turned it into a reference beam [15,16]. Then, lens L7 ($f$ = 75 mm) was used to Fourier transform the beam back into the image domain, while separating the two-wavelength beam into its two wavelength channels using dichroic mirror DM (cut-off wavelength 593 nm, FF593-Di03, Semrock). Each of the channels was then back-reflected by slightly tilted mirrors, M2 and M3, and projected to the camera at an off-axis angle through lenses L7 and L8. The relative angle between mirrors M2 and M3 was adjusted such that each wavelength channel created orthogonal off-axis interference fringe direction with respect to the other channel, and thus a single multiplexed off-axis hologram from the two wavelength channels can be recorded by the camera at a single exposure. As shown at the bottom inset of Fig. 1(b), the beams reflected from mirrors M2 and M3 would have hit the pinhole plate inside its solid disk area, and to avoid this, two off-axis holes were drilled into the pinhole disk, which allowed passage of the back-reflected reference beams to the camera. Note that by having the same amount of lenses in optical paths the sample beam and reference beams, it was possible to create beam-path and beam-curvature matchings between the beams within the coherence length of the illumination source used. Also note that in contrast to Ref. [15], implementing field of view multiplexing in the sample beam, the off-axis reference beams in the current letter, implementing wavelength multiplexing, are reflected back in the image domain, rather than in the Fourier domain, and thus 90°-rotated titled mirrors are used here rather than retro-reflectors.

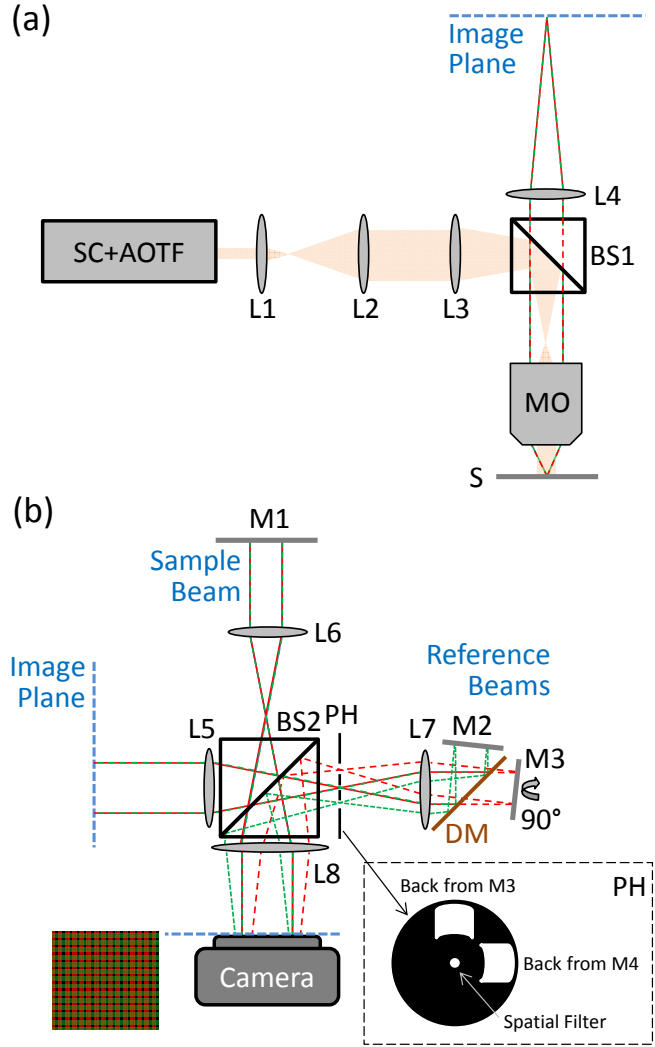

**Fig. 1.** (a) Dual-wavelength reflectance microscopy setup used for the experimental demonstrations. (b) External dual-wavelength off-axis multiplexed interferometric module for two-wavelength phase unwrapping. SC, Supercontinuum source. AOTF, Acousto-optical tunable filter. L1-L8, achromatic lenses. BS1, BS2, beam splitters, MO, Microscope objective. S, Sample. PH, Pinhole plate. DM, Dichroic mirror. Camera, Monochrome digital camera.

Assuming that the first wavelength channel induces straight interference fringes across the $x$ axis and that the second wavelength channel induces straight interference fingers across the $y$ axis, the multiplexed dual-wavelength interferogram, acquired by the monochrome camera in a single exposure is shown in Fig. 2(a), and can be expressed as follows:

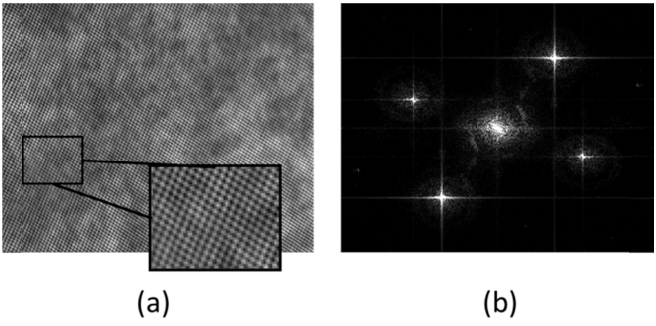

**Fig. 2.** (a) The multiplexed dual-wavelength interferogram, acquired in a single exposure. (b) The power spectrum of the multiplexed interferogram, as obtained by a digital 2D Fourier transform.

$$I(x,y) = I_{S1} + I_{R1} + \sqrt{I_{S1}I_{R1}}\cos[\phi_1(x,y) + x \cdot k_1 \sin(\alpha_1)] \quad (1)$$
$$+ I_{S2} + I_{R2} + \sqrt{I_{S2}I_{R2}}\cos[\phi_2(x,y) + y \cdot k_2 \sin(\alpha_2)],$$

where $I_{S1}$ and $I_{R1}$ are the sample and reference beam intensities, respectively, $\phi_1$ is the phase difference, and $k_1 \cdot \sin(\alpha_1)$ is the modulation term, comprised of wave number $k_1$ and the sine of the off-axis angle $\alpha_1$ between the sample and the reference beams, all for the first wavelength channel, and the same signs are valid for the second wavelength channel as well, but with subscript of 2 instead of 1.

The acquired multiplexed interferogram is 2D Fourier transformed digitally, and the power spectrum of the result is shown in Fig. 2(b). Since the multiplexed interferogram is composed of two orthogonal fringe directions, each of which belongs to another wavelength channel, the spectrum contains two pairs of fully separable cross-correlation terms. One term from each of the pairs can be cropped and processed according to the conventional off-axis holography phase extraction [4], but on each cross-correlation term separately. Shortly, after an inverse 2D Fourier transform on the selected cross-correlation term, the associated sample complex wave-front is obtained. The angle argument of this complex matrix is the wrapped phase of the associated wavelength channel. Phase wrapping occurs due to the $2\pi$ periodic nature of the arctan function, which is used to extract the angle of the complex wave-front, and it restricts the optical thickness of the sample to be within one wavelength, in order to be directly deciphered. For continuous and smooth thickness objects, digital unwrapping algorithms can be applied to unwrap the phase [5]. However, actual sharp phase discontinuities of over one wavelength would result in an error, or wrong interpretation of the phase map.

The two-wavelength phase retrieval method simplifies the digital process required and broadens the unambiguous phase range by processing the two phase profiles and creating another phase profile corresponding to a synthetic wavelength, which is much longer than any of the experimental wavelengths used [7].

In our case, for round-trip interaction with the sample, the extracted wrapped phases of each channel (marked with subscript 1 or 2) can be described as:

$$\phi_{1,2}(x,y) = 2 \cdot h(x,y) \cdot 2\pi \cdot \frac{1}{\lambda_{1,2}}, \quad (2)$$

where $h(x,y)$ denotes the optical thickness of the sample and is extracted as follows:

$$\phi_\Lambda(x,y) = \phi_1(x,y) - \phi_2(x,y) = 2 \cdot h(x,y) \cdot 2\pi \cdot \left(\frac{\lambda_2 - \lambda_1}{\lambda_1 \lambda_2}\right), \quad (3)$$
$$h(x,y) = \frac{\phi_\Lambda(x,y)}{4\pi}\left(\frac{\lambda_1 \lambda_2}{\lambda_2 - \lambda_1}\right) = \frac{\phi_\Lambda(x,y)}{4\pi}\Lambda,$$

where the synthetic wavelength is given by $\Lambda = \lambda_1 \cdot \lambda_2 / (\lambda_1 - \lambda_2)$. For instance, in our first case, where the central wavelengths are $\lambda_1 =$ 580 nm and $\lambda_2 =$ 597 nm, we obtained $\Lambda =$ 2,036 nm. Following this subtraction, some phase jumps may be present due to wrapping spatial mismatch between both wrapped phase maps, which are resolved by adding $2\pi$ when the difference is negative [7].

In general, by using spectrally closer wavelengths, $\Lambda$ increases, allowing a broader unambiguous range, thus thicker objects to be imaged without phase unwrapping. Two problems prevent us from using closer spectral channels. The first is related to the nature of reflection-based interferometry. Interferometric systems with high-coherence illumination are usually easier to align. However, when considering reflection-based interferometry system, any back-reflection from the various optical components in the system may result in parasitic interference patterns, which increases noise. Therefore, it is crucial to use spectrally broader illumination that still allows obtaining interference, but produces a shorter coherence length. This immediately dictates a minimal spectral distance between both wavelength channels. The second problem is inherent to dual-wavelength interferometry in general. As described elsewhere [7], the technique does not only increase the synthetic wavelength, but also increases noise levels in a relation to that of $\Lambda$ and $\lambda_{1,2}$. For instance, in our case, the phase noise was ~35 times greater in the new synthetic-wavelength phase map.

The dual-wavelength multiplexed interferometric module was first tested for accuracy and precision by imaging a 7.96 μm step test target (SHS, Model #VDS-8.0QS, Bruker). Several different areas on the step were captured, and each measurement was repeated 20 times for repeatability analysis. Each dual-wavelength multiplexed interferogram was processed as described above. The nominal step value was 7.96 μm, as measured by commercial white-light interferometry (WLI) profiler (Contour GT, Bruker).

After imaging the step target with the proposed interferometric module in a single exposure, we extracted the wrapped phase profiles associated with each of the two-wavelength channels, the cross-sections of which are shown in Fig. 3(a). As seen in this figure, there is a steep jump at the step position, which cannot be simply solved with digital unwrapping algorithms, as verified on each phase profile separately. However, after applying the two-wavelength unwrapping method, we obtained the phase profile cross-section shown by the blue curve in Fig. 3(b), where the previous cross-sections of the wrapped phase are shown at the bottom part of this figure, for reference.

Figure 3(c) shows the average values and the standard deviations of the two Gaussians curves obtained in the histogram of the height map. The step height was calculated as 7.92 μm from the difference of both averages. The average accuracy of the step height measurement (in relation to the reference WLI measurement) was between 10 and 60 nm, and the largest value in the system repeatability test (standard deviation over 20 frames) was 40 nm. As mentioned, spatial noise is amplified due to using synthetic wavelength, and in our case reached up to a maximum of 500 nm. This increased noise level problem due to the synthetic wavelength can be solved by algorithmic means [7]. Figure 3(d) shows 3D rendering of the step test-target height map.

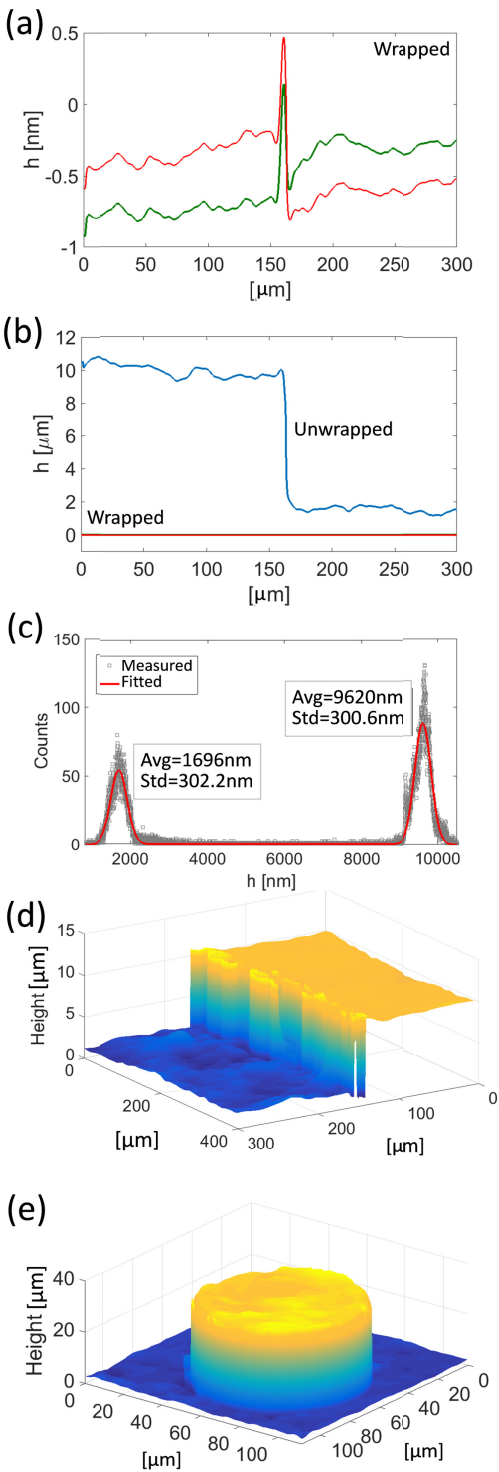

**Fig. 3.** Simultaneous two-wavelength phase unwrapping results obtained by the dual-wavelength module. (a) Two-wavelength phase profile cross-sections for a 7.96-μm-high step target. (b) The unwrapped phase cross-section, as obtained for the step target by using two-wavelength phase unwrapping. (c) A histogram of the height measurements for the step target. (d) A height-map 3D rendering for the step target. (d) A height-map 3D rendering for a 30.5-μm-high copper pillar.

In order to further explore applicative feasibility of the system for optical metrology, a round copper pillars sample was measured. The pillars nominal height was 30.5 μm and their diameter was 70 μm. In this case, we wanted to measure the actual height and its deviation from the nominal value, with a certainty of 7 μm. Thus, we chose $\lambda_1$ and $\lambda_2$ as 580 nm and 605 nm, respectively, producing a synthetic wavelength of 14.036 μm (approximately double than 7 μm). Since the nominal height of the pillars was known, and was larger than the depth of field of the microscope objective, we added to the masked pillars 4 full multiplications of $\lambda/2$ to reach an unambiguous range of 7 μm around $h = 28$ μm. Figure 3(e) shows 3D rendering of the pillar-target height map. Overall, 26 pillars were measured, with an average height of 30.59 μm, and a standard deviation of 0.56 μm. The average distance from the nominal height was 0.45μm.

To conclude, we presented a dual-wavelength multiplexed interferometry external module for simultaneous two-wavelength unwrapping. The module can be used as a modular add-on to existing microscopic systems, and enables two-wavelength simultaneous imaging of objects, which are optically thicker than the illumination wavelength. The module performs optical multiplexing of off-axis holograms, while both holograms share the same dynamic range of the sensor with negligible effects on the measurement accuracy. The accuracy obtained was better than 60 nm and repeatability was lower than 40 nm, over 20 measurements. These results are at the grade of commercial systems, and thus the proposed module might find uses in metrology and 3D micro-topography imaging applications, especially for dynamic samples, when two-wavelength channels need to be acquired simultaneously.

**Funding.** European Research Commission (ERC) (678316).

**References with titles**